\documentclass{PoS}
\usepackage{amsmath}
\usepackage{amssymb}
\usepackage{booktabs}
\usepackage{epstopdf}
\usepackage[utf8x]{inputenc}

\title{Implementation of the Neuberger-Dirac operator on GPUs}

\ShortTitle{The Neuberger-Dirac operator on the GPU}

\author{\speaker{Bjoern Walk} and Hartmut Wittig\\
        Institut für Kernphysik, Johannes Gutenberg-Universität Mainz,\\
        Johann Joachim Becher-Weg 45, 55099 Mainz, Germany\\
        E-mail: \email{bwalk@kph.uni-mainz.de,wittig@kph.uni-mainz.de}}
\author{Egor Dranischnikow and Elmar Schömer\\
        Institut für Informatik, Johannes Gutenberg-Universität Mainz,\\
        Staudingerweg 9, 55099 Mainz, Germany\\
        E-mail:
        \email{\{dranisch,schoemer\}@informatik.uni-mainz.de}}

\abstract{
  Recent developments have shown that a lot can be gained for QCD simulations
  from GPU hardware. This can be exploited especially in the case of
  Ginsparg-Wilson fermions when the computational costs are particularly high.
  In this work, we use the Neuberger-Dirac operator as our realisation of
  Ginsparg-Wilson fermions, which greatly facilitate lattice investigations of
  decays like $K \to \pi\pi$. We report on the ongoing study of our GPU
  implementation of the Neuberger-Dirac operator including the exact treatment
  of the low lying eigenmodes of the Wilson-Dirac operator. Our benchmarks
  show that we achieve speed-up factors of around 23 and 16 in single and
  double precision, respectively.
}

\FullConference{The XXVIII International Symposium on Lattice Field Theory, Lattice2010\\
		June 14-19, 2010\\
		Villasimius, Italy}

\DeclareMathOperator{\sign}{sign}
\newcommand{\PP}{\mathbb{P}}

\begin{document}

\section{Introduction}
In the past years the GPU, the core of hardware responsible to display images
on computer screens, became more and more flexible and powerful. The GPU soon
surpassed the CPU in terms of raw computational power, and various fields in
science found a way to exploit this (see also the review in
\cite{Clark:2009qp}). For the lattice community, ref.~\cite{Egri:2006zm} set
out one ground-breaking paper in which it was demonstrated how to map the
Wilson-Dirac operator on the GPU. With the introduction of NVIDIA CUDA, a
flexible programming model was given and made it even easier to access the
ever-increasing computational power. This was the basis for the reference
paper \cite{Barros:2008rd}.

The GPU approach could be particularly important when it comes to simulations
with Ginsparg-Wilson fermions \cite{GW} which possess an order-of-magnitude
higher computational costs than Wilson fermions.

One particular solution to the Ginsparg-Wilson relation is the Neuberger-Dirac
operator which we are studying in this paper. A lattice Dirac operator which
satisfies chiral symmetry at nonzero lattice spacing greatly facilitates the
investigation of certain processes like non-leptonic kaon decays. These decays
raise the long-standing problem for QCD phenomenology why the $\Delta I = 1/2$
amplitudes are so much larger than the $\Delta I = 3/2$ amplitudes. The bulk
of the enhancement must be due to strong interactions \cite{GLAM74} at low
energies and therefore, a reliable explanation must be based on systematic
non-perturbative methods, in particular on lattice QCD \cite{CB84}.

So far, a lot of effort has been invested to understand and quantify the
effects responsible for the enhancement of the $\Delta I=1/2$ amplitudes. A
series of simulations were performed using quenched calculations in the
SU(4)-symmetric case in order to understand the effect of the charm quark
\cite{RW}. One particular approach is to do simulations outside the GIM-limit
with an active charm quark and reasonable large volumes to study finite-volume
effect.

In this paper we report on the ongoing development of our tools for numerical
simulations in the $\epsilon$-regime utilizing GPU hardware for acceleration.
We show details of the theoretical background and implementation for our
Wilson-Dirac kernel and for the Neuberger-Dirac operator. We give performance
results for both operators and the calculation of low lying eigenmodes of the
Wilson-Dirac operator. At the end we give an outlook about planned work.

\section{Implementation details}
\subsection{The Wilson-Dirac operator}
Following Wilsons formulation for lattice QCD, the $\gamma_5$-Hermitian,
massive Wilson-Dirac operator $Q$ used in our implementation is defined by
\begin{align}
   D_\text{W} \phi(x) = (4 + m) \phi(x) -
    \frac{1}{2} \sum_{\mu = \pm 0}^{\pm 3} U_\mu(x) (1 - \gamma_{\mu})
    \phi(x + a \hat\mu),
    \quad Q = \gamma_5 D_\text{W},
\end{align}
where $m$ is the bare mass of the fermion. Gauge links $U_\mu(x)$ and the
Dirac matrices $\gamma_\mu$ follow the definition
\begin{align}
  U_{-\mu}(x) = U^{-1}_\mu (x - a \hat\mu) \quad \text{and} \quad
  \gamma_{-\mu} = -\gamma_\mu.
\end{align}

The most important part in the implementation of the operator is the data
layout. Data read and written to global device memory on the GPU has to
fulfill several constraints, known as coalescing rules \cite{NV}, to ensure
maximum performance. One of them requires that consecutive threads read data
words in sequential order. A data word can have 32-, 64- or 128-Bytes. In
\cite{Egri:2006zm} a data layout for the fermions fields and the gauge links
was introduced which is nowadays considered as the default and our
implementation follows that, too.

Since the Wilson-Dirac kernel is memory-bound, we intend to minimize memory
access. It is possible to save memory load instructions by realizing that
one does not need the whole 18 entries of the gauge link matrices as they
fulfill the orthogonalization relation $U^\dagger U = 1$ of the SU(3) gauge
group. In total, we need a minimum of 8 parameters to uniquely define an
element of the gauge links. Each element of SU(3) can be expressed as a linear
combination of Gell-Mann-matrices, and in principle one could choose the
coefficients in this linear combination as parameters. However, although the
reconstruction comes practically for free, we should not introduce any
overhead in the reconstruction. A reconstruction based on Gell-Mann matrices
is not beneficial in terms of numerical effort. A much better approach is
presented in \cite{Barros:2008rd,Bunk:1985rg} which also has the advantage,
that the inverse operation, i.e.\ finding the parameters for a given
SU(3)-matrix, is simple. Another simple reconstruction scheme is based on the
property
\begin{align}
  U = \begin{pmatrix}
    a_1 & b_1 & c_1 \\
    a_2 & b_2 & c_2 \\
    a_3 & b_3 & c_3
  \end{pmatrix} \qquad
  \vec{c} = (\vec{a} \times \vec{b})^*,
\end{align}
that one column of the matrix can be reconstructed by the other two.

The rest of the implementation is fairly straightforward since there is no
communication necessary if we assign each lattice site of the resulting
fermion field a single thread. We explicitly write down the multiplication
in Dirac-space of the eight directions $\mu = \pm 0 \ldots \pm 3$, unroll the
matrix-vector multiplication in color space after reconstruction and
accumulate the resulting fermion in local registers. After the multiplication
with $\gamma_5$, we write the result to global device memory.

\subsection{The Neuberger-Dirac operator}
Following the conventions and notations from \cite{GHLW02}, the
Neuberger-Dirac operator $D_\text{N}$ \cite{Neuberger:1997fp} can be defined
in terms of the Wilson-Dirac operator $D_\text{W}$
\begin{align}
  D_\text{N} = \frac{1 + \gamma_5 \, \sign(Q)}{\bar a},
  \qquad Q = \gamma_5 (a D_\text{W} - 1 - s).
\end{align}
In this definition, $|s| < 1$ is a tunable parameter while the $\sign$-function
has to be defined by its series expansion. For numerical stability we choose
Chebychev polynomials, so the $\sign$-function of the operator $Q$ is given by
\begin{align}
  \sign (Q) \simeq X P_n(X^2),
  \qquad X \equiv \frac{Q}{||Q||}
  \quad \text{and} \quad P_n(x) = \sum_{k=0}^n c_k T_k(x).
  \label{uniapprox}
\end{align}
The definition of the Chebychev polynomials $T_k(y)$ can be found in
\cite{NR}. Because Chebychev polynomials fulfill a recursion relation they
permit an evaluation via the Clenshaw recurrence formula.

In order to find the coefficients $c_k$ of the expansion, we aim to minimize
the error
\begin{align}
  \delta = \max_{\epsilon \leq y \leq 1} \vert h(y) \vert,
  \qquad h(y) \equiv 1 - \sqrt{y} P(y)
\end{align}
of the polynomial expansion for specified $\epsilon$. In the range
$\sqrt{\epsilon} \leq \vert x \vert \leq 1$ the function $x P(x^2)$ then
approximates $\sign(x)$ uniformly with a maximal deviation $\delta$.
Polynomials constructed in such a way are often referred to as minmax
polynomials. If $\epsilon$ is chosen such that $Q^2 \geq \epsilon \Vert Q
\Vert^2$, the error in (\ref{uniapprox}) is an operator with norm less than or
equal to $\delta$. The approximation error is always bounded by $\delta \Vert
\eta \Vert$, uniformly in the field $\eta$ to which the operator is applied.

This method is straightforward, but in the case of the $\epsilon$-regime
actually not recommended. In this case, the operator $Q^2$ may have some
exceptionally low-lying eigenvalues, and it is far more efficient first to
separate the few lowest modes and to treat them exactly.

\subsection{Low-mode projection}
The spectrum of $Q$ in the vicinity of the origin can be reliably determined
by minimizing the Ritz functional of $Q^2$ \cite{B94}. Thereby, we also find
an approximation of the associated eigenvectors. In order to control the error
of the total approximation, we need to examine by how much these vectors
deviate from the true eigenvector \cite{GHLW02}.

Assuming that a specified number $l$ of approximate eigenvectors has been
computed, we denote the linear space spanned by these vectors by $V$ and by
$\PP$ the corresponding orthonormal projector. We shall take it for granted
that the eigenvalues $\nu_k$ of $Q$ are separated from zero and from the rest
of the spectrum by a distance greater than $\rho$. This number measures the
deviation of V from being an exact eigenspace of Q. The presence of a spectral
gap around zero implies that the subset of positive and negative eigenvalues
can be identified without any numerical ambiguity. With the introduction of
the eigenvectors $u_k \in V$, $\PP Q u_k = \nu_k u_k$, the associated
orthonormal projectors are given by
\begin{align}
  \PP_+ = \sum_{\nu_k > 0} u_k \otimes (u_k)^\dagger
  \quad \text{and} \quad
  \PP_- = \sum_{\nu_k < 0} u_k \otimes (u_k)^\dagger
\end{align}
respectively. In the same way we may define the projectors
$(\PP_\pm)_\text{exact}$ to the subspaces spanned by the corresponding exact
eigenvectors of $Q$ with positive and negative eigenvalues $\lambda_k$. We can
give a quantitative estimate on the deviation of the computed projectors
$\PP_\pm$ from the exact projectors $(\PP_\pm)_\text{exact}$. The total error
depends on the size of the residues
\begin{align}
  \rho_k = \Vert (Q - \nu_k) u_k \Vert
\end{align}
and also on the distance between the eigenvalues, as a small value of $\rho_k$
does not exclude sizeable mixing of $u_k$ with several eigenvectors of $Q$
if these have eigenvalues that are within a distance $\rho_k$ of $\nu_k$. When
estimating the deviation of the projectors rather than that of the individual
eigenvectors, the interesting quantity is the distance $d_k$ of $\nu_k$ from
the exact spectrum of $Q$ in the subspace that is orthogonal to the range of
$(\PP_+)_\text{exact}$ if $\nu_k > 0$ if $(\PP_-)_\text{exact}$ if $\nu_k <
0$. The quality of the approximation is then controlled by the parameters
\begin{align}
  \kappa_\pm^2 = \sum_{\pm \nu_k > 0} \rho_k^2 / d_k^2,
  \quad \kappa_\pm > 0
\end{align}
and an upper bound for the deviation of the projectors can by given by
\begin{align}
  \Vert \PP_\pm - (\PP_\pm)_\text{exact} \Vert \leq\
  \frac{\kappa_\pm (1 + 2 \kappa_\pm)}{1 - 2 \kappa_\pm (1 + 2 \kappa_\pm)}.
\end{align} 

In our implementation, the whole computation of the projectors is performed on
the GPU. The number of low modes included in the projector will be determined
dynamically in such a way that the spectral distance from the other modes is
not accidentally small. The parameters $\kappa_\pm$ can be estimated without
difficulty and the Ritz functional is stopped when the desired level of
precision is reached. Afterwards, we replace Equation~\ref{uniapprox} by
\begin{align}
  \sign(Q) \simeq \PP_+ - \PP_- + (1 - \PP_+ - \PP_-) X P_n(X^2),
\end{align}
and it can be shown that the approximation $\tilde D_\text{N}^m$ to the
massive Neuberger-Dirac operator satisfies
\begin{align}
  \Vert \tilde D_\text{N}^m - D_\text{N}^m \Vert \leq
  \frac{2}{a} \bigl( 1 + s - \frac{am}{2} \bigr) \bigl( \kappa_+ + \kappa_-
  \bigr).
\end{align}

\section{Results}

For performance results we have tested the GeForce GTX285 GPU, a current
generation card, and the GeForce GTX480, as well as a Tesla C2050 GPU, the
latter of which are both based on the recently released Fermi chipset. The
code, however, was not optimized for the new architecture and we expect
further enhancements in the future. An overview of the key features of each
GPU is given in Table~\ref{hardware}.

\begin{table}
  \centering
  \begin{tabular}{rccc}
    \toprule
    & GTX285 & GTX480 & C2050 \\
    \midrule
    No. of Cores & 240 & 480 & 448 \\
    Memory amount [MB] & 2048 & 1536 & 3072 \\
    Shader Clock [MHz] & 1476 & 1401 & 1150 \\
    Memory Clock [MHz] & 1242 & 1848 & 1500 \\
    Memory Bandwidth [GB/s] & 159.0 & 177.4 & 144.0 \\
    \bottomrule
  \end{tabular}
  \caption{\label{hardware}
    Key specification of the three NVIDIA GPUs available for benchmarking. For
    the C2050, roughly 10\% of total memory amount has to be subtracted if
    the error correction ECC is enabled by the driver.
  }
\end{table}

We have already stated that the performance of the Wilson-Dirac kernel is
memory bound. In our optimization we aimed for maximizing the total memory
bandwidth achieved by our implementation. This also means that we should not
expect strong scaling as the number of cores is increased. The major
contribution to the performance of the calculation comes from the memory
bandwidth.

\begin{figure}
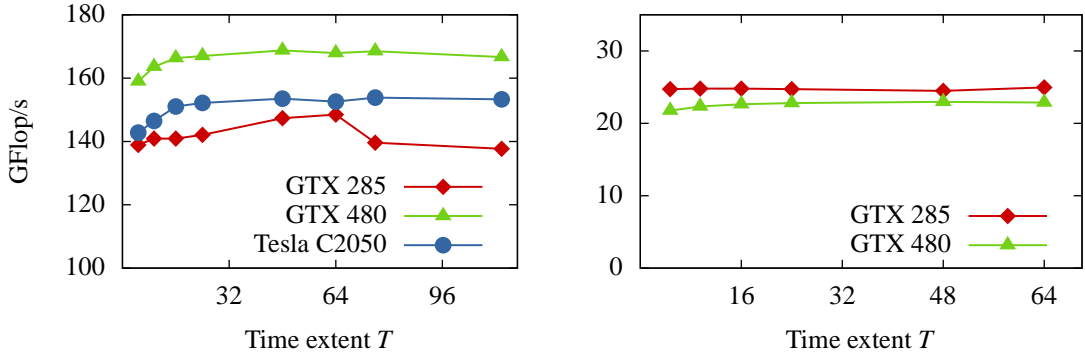

  \centering
  \begin{minipage}{.49\textwidth}
    \small{\input{fig-qphi.tex}}
  \end{minipage}
  \begin{minipage}{.49\textwidth}
    \small{\input{fig-qphi64.tex}}
  \end{minipage}
  \caption{\label{qphi}
    Performance result for the application of the massive Wilson-Dirac
    operator $Q$. \textbf{LEFT:} Single precision performance. \textbf{RIGHT:}
    Double precision performance. The spatial extent $L$ is fixed at $24^3$
    throughout.
  }
\end{figure}

In Figure~\ref{qphi} the performance of the Wilson-Dirac kernel in single- and
double precision is given as a function of the lattice volume. The
double precision data curve on the C2050 is missing because we had only
limited access to the GPU and it was not available anymore at the time of our
testings.

We can see a nearly constant behaviour for sufficient lattice volumes. On the
GTX285 we observe a small peak at $T = 64$ which we believe is largely
accidental. For double precision we see a decrease of the performance by a
factor of around 8. On the GTX285, this comes from the fact that there is only
1 double precision unit while 8 single precision units are available. In
principle, the architectural design of the GTX480 should give up to half of
the single precision speed in double precision. NVIDIA, however, restricts
this to a quarter of the single precision speed inside the driver. This drop
in performance can be explained by the fact that we did not optimize for the
new chipset. The final result for the Wilson-Dirac kernel gives around $140 \,
\text{GFlop/s}$ on the GTX285, $150 \, \text{GFlop/s}$ on the C2050 and $170
\, \text{GFlop/s}$ on the GTX480 in single precision. For double precision we
achieve $22 \, \text{GFlop/s}$ and $25 \, \text{GFlop/s}$ on the GTX285 and
GTX480 respectively.

\begin{figure}
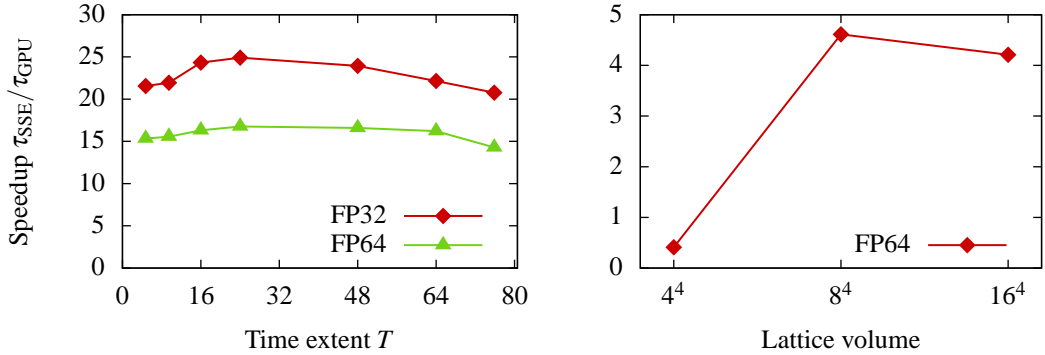

  \centering
  \begin{minipage}{.49\textwidth}
    \small{\input{fig-dphi.tex}}
  \end{minipage}
  \begin{minipage}{.49\textwidth}
    \small{\input{fig-lm.tex}}
  \end{minipage}
  \caption{\label{dphi}
    \textbf{LEFT:} Performance result of the application of the
    Neuberger-Dirac operator $D$ as compared to the SSE2-optimized CPU
    version. The spatial extend $L$ is fixed at $24^3$.
    \textbf{RIGHT:} Performance result for the calculation of low-lying
    eigenmodes of the Wilson-Dirac operator $Q$.
    The plots show the speed-up factor $\tau_\text{SSE}/\tau_\text{GPU}$ in
    single and double precision, respectively. 
  }
\end{figure}

In Figure~\ref{dphi} the performance of the Neuberger-Dirac operator in
single and double precision is illustrated as a comparison to the
SSE2-optimized CPU version. The execution time of the operator is normalized
on the number of Clenshaw iterations and the lattice volume. The CPU version
was run on an Intel E6400 Core2 with a single core clock rate of $2.13 \,
\text{GHz}$. The GPU version was run on the GTX285. We can observe a nearly
constant behaviour in the execution time in dependence to the lattice volume.
For single precision, the SSE2 version has an average of $\tau_\text{SSE} =
0.877 \, \mu\text{s}$ and the GPU version of $\tau_\text{GPU} = 0.038 \,
\mu\text{s}$, giving a speedup factor of around 23. For double precision, the
numbers read $\tau_\text{SSE} = 1.459 \, \mu\text{s}$ and $\tau_\text{GPU} =
0.090 \, \mu\text{s}$, hence, a speedup factor of around 16.

\section{Conclusions and outlook}
In this proceedings article, we gave a first introduction of our study of a
GPU-based simulation program for lattice QCD based on the Neuberger-Dirac
operator. We introduced the basic theoretical background and gave first
performance results of our implementation of the Wilson-Dirac operator and the
Neuberger-Dirac operator. We have shown that the performance for the
Wilson-Dirac operator is of the same order of magnitude compared to other
implementations previously published. For the Neuberger-Dirac operator, we
have shown that we can reach a speedup factor to the Wilson-Dirac operator of
around 23 on single precision and 16 on double precision. We are going to
develop code for the calculation of the index of our Neuberger-Dirac operator
via zero-mode counting \cite{RW}. In the long run, we aim to integrate our
code into a larger set of simulation programs to calculate observables in the
process $K \to \pi\pi$ in order to increase the lattice volume on those
calculations.

\section{Acknowledgements}
Part of the performance results were obtained on GPU of the KOMET
collaboration of the University of Mainz. We are indebted to the institute
for these opportunity. B.~W. is funded by the DFG via GK 1581.


\begin{thebibliography}{99}
\bibitem{Clark:2009qp}
  M.~A.~Clark,
  %``QCD on GPUs: cost effective supercomputing,''
  PoS {\bf LATTICE2009} (2009) 003
  [arXiv:0912.2268 [hep-lat]].
  %%CITATION = POSCI,LATTICE2009,003;%%
\bibitem{Egri:2006zm}
  G.~I.~Egri, Z.~Fodor, C.~Hoelbling, S.~D.~Katz, D.~Nogradi and K.~K.~Szabo,
  %``Lattice QCD as a video game,''
  Comput.\ Phys.\ Commun.\  {\bf 177} (2007) 631.
  %%CITATION = CPHCB,177,631;%%
\bibitem{Barros:2008rd}
  K.~Barros, R.~Babich, R.~Brower, M.~A.~Clark and C.~Rebbi,
  PoS {\bf LATTICE2008} (2008) 045.
\bibitem{GW}
  P.~H.~Ginsparg and K.~G.~Wilson,
  %``A Remnant Of Chiral Symmetry On The Lattice,''
  Phys.\ Rev.\  D {\bf 25} (1982) 2649;
  %%CITATION = PHRVA,D25,2649;%%
  D.~B.~Kaplan,
  %``Chiral fermions on the lattice,''
  Nucl.\ Phys.\ Proc.\ Suppl.\  {\bf 30} (1993) 597;
  %%CITATION = NUPHZ,30,597;%%
  R.~Narayanan and H.~Neuberger,
  %``A Construction of lattice chiral gauge theories,''
  Nucl.\ Phys.\  B {\bf 443}, 305 (1995).
  %%CITATION = NUPHA,B443,305;%%
\bibitem{GLAM74}
  M.~K.~Gaillard and B.~W.~Lee,
  %``Delta I = 1/2 Rule For Nonleptonic Decays In Asymptotically Free Field
  %Theories,''
  Phys.\ Rev.\ Lett.\  {\bf 33}, 108 (1974);
  %%CITATION = PRLTA,33,108;%%
  G.~Altarelli and L.~Maiani,
  %``Octet Enhancement Of Nonleptonic Weak Interactions In Asymptotically Free
  %Gauge Theories,''
  Phys.\ Lett.\  B {\bf 52}, 351 (1974).
  %%CITATION = PHLTA,B52,351;%%
\bibitem{CB84}
  N.~Cabibbo, G.~Martinelli and R.~Petronzio,
  %``Weak Interactions On The Lattice,''
  Nucl.\ Phys.\  B {\bf 244}, 381 (1984);
  %%CITATION = NUPHA,B244,381;%%
  R.~C.~Brower, G.~Maturana, M.~Belen Gavela and R.~Gupta,
  %``Calculation Of Weak Transitions In Lattice QCD,''
  Phys.\ Rev.\ Lett.\  {\bf 53} (1984) 1318.
  %%CITATION = PRLTA,53,1318;%%
\bibitem{RW}
  P.~Hernandez, M.~Laine, C.~Pena, E.~Torro, J.~Wennekers and H.~Wittig,
  %``Determination of the $\Delta S = 1$ weak Hamiltonian in the SU(4) chiral
  %limit through topological zero-mode wave functions,''
  JHEP {\bf 0805}, 043 (2008);
  %%CITATION = JHEPA,0805,043;%%
  L.~Giusti, P.~Hernandez, M.~Laine, C.~Pena, J.~Wennekers and H.~Wittig,
  %``On K --> pi pi amplitudes with a light charm quark,''
  Phys.\ Rev.\ Lett.\  {\bf 98}, 082003 (2007);
  %%CITATION = PRLTA,98,082003;%%
  L.~Giusti, P.~Hernandez, M.~Laine, P.~Weisz and H.~Wittig,
  %``A strategy to study the role of the charm quark in explaining the  Delta(I)
  %= 1/2 rule,''
  JHEP {\bf 0411} (2004) 016;
  %%CITATION = JHEPA,0411,016;%%
  P.~Hernandez and M.~Laine,
  %``Charm mass dependence of the weak Hamiltonian in chiral perturbation
  %theory,''
  JHEP {\bf 0409} (2004) 018.
  %%CITATION = JHEPA,0409,018;%%
\bibitem{NV}
  NVIDIA,
  NVIDIA CUDA C Programming Guide.
\bibitem{Bunk:1985rg}
  B.~Bunk and R.~Sommer,
  %``An Eight Parameter Representation Of SU(3) Matrices And Its Application For
  %Simulating Lattice QCD,''
  Comput.\ Phys.\ Commun.\  {\bf 40} (1986) 229.
  %%CITATION = CPHCB,40,229;%%
\bibitem{GHLW02}
  L.~Giusti, C.~Hoelbling, M.~Lüscher and H.~Wittig,
  Commput.\ Phys.\ Commun.\ {\bf153} (1998) 31.
\bibitem{Neuberger:1997fp}
  H.~Neuberger,
  %``Exactly massless quarks on the lattice,''
  Phys.\ Lett.\  B {\bf 417} (1998) 141;
  %%CITATION = PHLTA,B417,141;%%
  H.~Neuberger,
  %``More about exactly massless quarks on the lattice,''
  Phys.\ Lett.\  B {\bf 427} (1998) 353.
  %%CITATION = PHLTA,B427,353;%%
\bibitem{NR}
  W.H.~Press, S.A.~Teukolsky, W.T.~Vetterling and B.P.~Flannery,
  Numerical Recipes, Third Edition.
\bibitem{B94}
  B.~Bunk, K.~Jansen, M.~Lüscher, H.~Simma,
  ALPHA collaboration internal report (1994);
  T.~Kalkreuther and H.~Simma,
  Comput.\ Phys.\ Comm.\ {\bf 93} (1996) 33.
\end{thebibliography}
\end{document}